\documentclass[proceedings]{JHEP3}
\usepackage{amsfonts}
\usepackage{amsmath}
\usepackage{epsfig,multicol}

\newbox\mybox

\newcommand\fverb{\setbox\mybox=\hbox\bgroup\verb}
\newcommand\fverbdo{\egroup\medskip\noindent\fbox{\unhbox\mybox}\ }
\newcommand\fverbit{\egroup\item[\fbox{\unhbox\mybox}]}
\conference{Particles versus fields}
\abstract{We review some recent results on how ${\mathcal{PT}}$-symmetry, that
is a simultaneous time-reversal and parity transformation, can be used to
construct new integrable models. Some complex valued multi-particle
systems, such as deformations of the Calogero-Moser-Sutherland models, are shown
to arise naturally from real valued field equations of non-linear integrable systems. Deformations of
complex non-linear integrable field equations, some of them even allowing
for compacton solutions, are also investigated.  The integrabilty of various
systems is established by means of the Painlev\'{e} test.}

\title{Particles versus fields in ${\mathcal{PT}}$-symmetrically deformed integrable systems}
\author{Andreas Fring \\
Centre for Mathematical Science, City University London, \\
Northampton Square, London EC1V 0HB, UK\\
E-mail: A.Fring@city.ac.uk}

\input{tcilatex}

\begin{document}

\section{Introduction}

There are many examples of non-Hermitian integrable systems in the literature pre-dating the
paper by Bender and Boettcher \cite{BB}, which gave rise to the recent wider interest in non-Hermitian
Hamiltonian systems. A well studied
class of quantum field theories is for instance affine Toda field theories (ATFT)
\begin{equation}
{\mathcal{L}} = \frac{1}{2}\partial _{\mu }\phi \partial ^{\mu }\phi +\frac{m^{2}}
{\beta ^{2}} \sum\limits_{k=0}^{\ell } n_{k}\exp (\beta \alpha _{k}\cdot
\phi), \label{Toda}
\end{equation}
involving $\ell$ scalar fields $\phi$. The $n_{k}$ are integers often called Kac labels and the $\alpha _{k}$ for
$k=1, \ldots, \ell $ are simple roots with $\alpha_0$ being the negative of the highest root.
When the coupling constant $\beta$ is taken to be purely imaginary these models
have interesting and richer features than their real counterparts. The classical solitons
were found \cite{Toda} to have real masses despite the fact that
the model is a non-Hermitian Hamiltonian system. Unlike as for real coupling, the
scattering of their fundamental particles allows for backscattering such that the associated Yang-Baxter equations give
rise to solutions in terms of representations of quasi-triangle Hopf algebras (quantum groups).
For the simplest example, the $A_1$-model, corresponding to complex Liouville theory,
a rigorous proof for the reality of the spectrum was found by Faddeev and Tirkkonen \cite{Fadd}
by relating it to the Hermitian XXZ-quantum spin chain using Bethe ansatz techniques.

In addition, integrable quantum spin chains of non-Hermitian type have been investigated in the past
for instance by von Gehlen \cite{Guenter}. The Ising quantum spin
chain in an imaginary field corresponds in the continuous limit to the Yang-Lee model ($A_2^{(2)}$-minimal ATFT)
\begin{equation}
{\mathcal{H}}=\frac{1}{2}\sum\limits_{i=1}^{N}\sigma _{i}^{x}+\lambda \sigma
_{i}^{z}\sigma _{i+1}^{z}+i \kappa \sigma _{i}^{z}, \qquad \qquad \lambda ,\kappa \in \mathbb{R},
\end{equation}
and may be used to describe phase transitions. Here $N$ denotes the length of the spin chain and
the $\sigma _{i}^{z}$, $\sigma _{i}^{x}$ are the usual Pauli matrices
describing spin $1/2$ particles and acting  on the site $i$ in the
state space of the form $({\mathbb{C}}^{2})^{\otimes N}$.

It is by now well understood how to explain the spectral properties of such models by
means of  ${\mathcal{PT}}$-symmetry, i.e. symmetry of the
Hamiltonian and the wavefunction with respect to a simultaneous parity transformation
and time reversal, pseudo-Hermiticity or quasi-Hermiticity \cite{Rev1,Rev2,Rev3}. In addition is also
established how to formulate a consistent quantum mechanical description via the
definition of a new metric, although this is only worked out in detail for very
few solvable models, e.g. \cite{PAAF}. Interesting questions regarding the uniqueness of the physical
observables still need further investigations and remain unanswered. Once of the new feature
is that unlike as for Hermitian systems the Hamiltonian alone is no longer
sufficient enough to define the set of observables \cite{Hendrik}.

For classical systems, which are the main subject of this article, one may also
use ${\mathcal{PT}}$-symmetry to establish the reality of the energy \cite{AFKdV}
\begin{equation}
E=\int\limits_{-a}^{a}{\mathcal{H}}[u(x)]dx=-\int\limits_{a}^{-a}{\mathcal{H
}}[u(-x)]dx=\int\limits_{-a}^{a}{\mathcal{H}}^{\dagger }[u(x)]dx=E^{\dagger}.  \label{clre}
\end{equation}
Note that unlike as for the quantum case the reality can be established from the Hamiltonian alone,
albeit together with some appropriate boundary conditions.

\subsection{${\mathcal{PT}}$-guided deformations} \label{def}
Let us now turn to the question of how to use the above mentioned arguments to
construct new consistent models with real spectra. In principle we could use any of them,
but clearly to exploit ${\mathcal{PT}}$-symmetry is most transparent, especially for classical models
as we indicated in (\ref{clre}). Keeping in mind that the effect of a ${\mathcal{PT}}$-transformation
is ${\mathcal{PT}}: x \rightarrow -x, p \rightarrow p$ and $i \rightarrow -i$, we may deform any ${\mathcal{PT}}$-symmetric
function in the following way
\begin{equation}
f(x) \rightarrow f[- i (ix)^{\varepsilon}], \;\;
                f[- i (ix)^\varepsilon p^{\varepsilon-1}], \; \; f(x) + \tilde{f}[(ix)^{\varepsilon}] + \hat{f}[(ix)^\varepsilon p^{\varepsilon}], \label{fdef}
\end{equation}
while keeping its invariance. The deformation parameter $\varepsilon \in {\mathbb{R}}$ is choosen in
such a way that the undeformed case is recovered for $\varepsilon=1$. The same principle may be applied
to derivatives of $\mathcal{PT}$-symmetric functions
\begin{equation}
             \partial _{x}f(x) \rightarrow f_{x;\varepsilon} :=
             \partial _{x,\varepsilon }f(x)=-i(if_{x})^{\varepsilon }, \; \; \partial _{x,\varepsilon }^{n}:=\partial _{x}^{n-1}\partial _{x,\varepsilon} ,
\end{equation}
such that
             \begin{eqnarray}
f_{xx;\varepsilon} :=\partial _{x,\varepsilon }^{2}f &=&-i\varepsilon (if_{x})^{\varepsilon }%
\frac{f_{xx}}{f_{x}}, \\
f_{xxx;\varepsilon} :=\partial _{x,\varepsilon }^{3}f &=&-i\varepsilon (if_{x})^{\varepsilon }
\left[ \frac{f_{xxx}}{f_{x}}+(\varepsilon -1)\left( \frac{f_{xx}}{f_{x}}%
\right) ^{2}\right],
\end{eqnarray}
and even to supersymmetric derivative of a $\mathcal{PT}$-symmetric functions
\begin{equation}
             D=\theta \partial _{x}+\partial _{\theta } \rightarrow D_{\varepsilon }:=\theta \partial _{x,\varepsilon }+\partial _{\theta },
\end{equation}
with $\theta$ being the usual anti-commuting superspace variable. Remarkably it can be shown that the latter deformation can be carried out without breaking the supersymmetry of the models  \cite{AFBijan}. We shall apply these deformations in section \ref{PTdef}

\section{From real fields to complex particle systems}

The above mentioned principle appears at times somewhat ad hoc and often the only motivation
provided is that such models are likely to have real spectra.
However, in the context of integrable systems some complex particle systems arise
very naturally when taking systems for real valued fields as starting points.

\subsection{No restrictions, $\ell $-soliton solution of the Benjamin-Ono
equation}

Let us consider a field equation for a real valued field $u(x,t)$ of the form
\begin{equation}
u(x,t)=\frac{\lambda }{2}\sum_{k=1}^{\ell }\left( \frac{i}{x-z_{k}(t)}-\frac{i}{
x-z_{k}^{\ast}(t)}\right), \qquad \lambda \in \mathbb{R}.  \label{re}
\end{equation}
Chen, Lee and Pereira showed thirty years ago \cite{Chen,Josh} that this Ansatz
constitutes an $\ell $-soliton solution for the
Benjamin-Ono equation \cite{BO}
\begin{equation}
u_{t}+uu_{x}+\lambda Hu_{xx}=0, \label{BO}
\end{equation}
with $Hu(x)$ denoting the Hilbert transform $Hu(x)=\frac{P}{\pi }\int_{-\infty
}^{\infty }\frac{u(x)}{z-x}dz,$ provided the $z_{k}$ in (\ref{re}) obey
the {\em complex}  $A_{\ell}$-Calogero equation of motion
\begin{equation}
\ddot{z}_{k}=\frac{\lambda ^{2}}{2}\sum\limits_{j\neq k}(z_{j}-z_{k})^{-3}, \qquad z_k \in {\mathbb{C}}.
\end{equation}
This is certainly the easiest example to demonstrate of how complex valued particle systems
arise naturally from real valued fields.

\subsection{Restriction to a submanifold}

Obviously we do not expect the above procedure to produce complex valued particle systems
when starting with any type of field equation. Dropping for instance in equation (\ref{BO})
the Hilbert transform and considering therefore Burgers equation instead will
not lead to the desired result. However, we may consistently impose some additional
constraints and make use of the following theorem found more than thirty years ago
by Airault, McKean and Moser \cite{AMM}:

\emph{Given a Hamiltonian $H(x_{1},\ldots ,x_{n},\dot{x}_{1},\ldots
,\dot{x}_{n})$ with flow
\begin{equation}
x_{i}={\partial H}/{\partial \dot{x}_{i}}\qquad \text{and\qquad }\ddot{x}%
_{i}=-{\partial H}/{\partial x_{i}}\qquad i=1,\ldots ,n
\end{equation}
and conserved charges $I_{j}$ in involution with $H$, i.e. vanishing Poisson brackets
$\{I_{j},H\}=0$. Then the locus of \texttt{grad} $I=0$ is invariant with regard to time evolution.
Thus it is permitted to restrict the flow to that locus provided it is not empty.}

In fact, often there are no real solutions to \texttt{grad} $I=0$ and one is once again
naturally led to consider complex particle systems. We consider the
Boussinesq equation, that is a set of coupled KdV type equations, as an example
\begin{equation}
v_{tt}=a(v^{2})_{xx}+bv_{xxxx}+v_{xx} \qquad a,b \in \mathbb{R} . \label{bou}
\end{equation}
Then the real valued field
\begin{equation}
v(x,t)=\lambda\sum\limits_{k=1}^{\ell } {(x-z_{k}(t))^{-2}}, \qquad \lambda \in \mathbb{R}
\end{equation}
satisfies the Boussinesq equation (\ref{bou}) if and only if $b=1/12$, $\lambda=-a/2$ and $z_{k}$ obeys
the constraining equations
\begin{eqnarray}
\ddot{z}_{k} &=&2\sum\limits_{j\neq k}(z_{j}-z_{k})^{-3}\qquad \quad \Leftrightarrow
\quad \ddot{z}_{k}=-\frac{\partial H_{Cal}}{\partial z_{i}}, \label{equom}\\
\dot{z}_{k}^2 &=&1-\sum\limits_{j\neq k}(z_{j}-z_{k})^{-2}\qquad \Leftrightarrow
\quad  \texttt{grad} (I_{3}-I_{1})=0. \label{con}
\end{eqnarray}
Here $I_{3}= \sum_{j=1}^{\ell} [ \dot{z}_{j}^3/3 + \sum\nolimits_{k\neq j} \dot{z}_{j} (z_{j}-z_{k})^{2}]$
and $I_{1}= \sum_{j=1}^{\ell} \dot{z}_{j}$ are two conserved charges in the $A_{\ell}$-Calogero model.
In principle it could be that there is no solution to these equations, meaning that the imposition 
of the additional constraint (\ref{con}), besides the equation of motion (\ref{equom}), will produce 
an empty locus. However, this is not the case and some genuine non-trivial solutions may be found. For $n=2$ a solution was
already reported in \cite{AMM}
\begin{equation}
z_1 = \kappa + \sqrt{(t+\tilde{\kappa})^2 +1/4}, \qquad  z_2 = \kappa - \sqrt{(t+\tilde{\kappa})^2 +1/4}
\end{equation}
such that the Boussinesq solution acquires the form
\begin{equation}
v(x,t)= 2 \lambda
\frac{(x-\protect\kappa )^{2}+(t+\tilde{\protect\kappa})^{2}+1/4}{[(x-%
\protect\kappa )^{2}-(t+\tilde{\protect\kappa})^{2}-1/4]^{2}}. \label{v}
\end{equation}
Note that $v(x,t)$ is still a real solution. Without any complication we may change $\kappa$
and $\tilde{\kappa}$ to be purely imaginary in which case, and only in this case, (\ref{v}) becomes a
solution for the ${\mathcal{PT}}$-symmetric equation (\ref{bou}) in the sense that
${\mathcal{PT}}: x\rightarrow -x, t\rightarrow -t$  and $v\rightarrow v$. Different
types of solutions and also for other values of $n$ will be reported elsewhere \cite{PAM}.

\section{${\mathcal{PT}}$-deformed particle systems} \label{PTdef}

Having presented some examples of how to obtain complex many particle
systems in a very natural way from real valued field equations, it
appears less ad hoc to start directly by deforming some integrable
many-body problems according to the principles described in section \ref{def},
having in mind that there might exist a corresponding real valued field equation.

\subsection{Complex extended Calogero-Moser-Sutherland (CMS) models}

The simplest way to deform a given model is just by adding a term to it along the
lines indicated in equation (\ref{fdef}). For a many body-system this was first proposed for the 
$A_\ell$-Calogero model in \cite{BK}
\begin{equation}
{\mathcal{H}}_{BK}=\frac{p^{2}}{2}+\frac{\omega ^{2}}{2}\sum
\limits_{i}q_{i}^{2}+\frac{g^{2}}{2}\sum\limits_{i\neq k}\frac{1}{
(q_{i}-q_{k})^{2}}+i\tilde{g}\sum\limits_{i\neq k}\frac{1}{(q_{i}-q_{k})}
p_{i}, \label{BK}
\end{equation}
with $g,\tilde{g}\in {\mathbb{R}},q,p\in {\mathbb{R}}^{\ell +1}$. The Hamiltonian
${\mathcal{H}}_{BK}$ differs from the usual Calogero model by the last term. There are
some immediate questions to be raised with regard to (\ref{BK}). Is it possible to have a
representation independent formulation for ${\mathcal{H}}_{BK}$? May one use other algebras or
Coxeter groups besides $A_\ell$ and  $B_\ell$? Is it possible to use non-rational potentials? Can one have
more coupling constants? Are the extensions still integrable? These questions were answered in
\cite{AF}, where it was noticed that one may generalize the Hamiltonian ${\mathcal{H}}_{BK}$ to
\begin{equation}
{\mathcal{H}}_{\mu }=\frac{1}{2}p^{2}+\frac{1}{2}\sum\limits_{\alpha \in
\Delta }g_{\alpha }^{2}V(\alpha \cdot q)+i\mu \cdot p, \label{af}
\end{equation}
with $\Delta $ being any root system and the new vector $\mu =1/2\sum\limits_{\alpha \in \Delta }\tilde{g}_{\alpha }f(\alpha \cdot q)\alpha $, with $f(x)=1/x$ and $V(x)=f^{2}(x)$. It is not so obvious, in fact no case independent proof
 is known, that one can further re-write the Hamiltonian such that it becomes the standard Hermitian Calogero Hamiltonian
with shifted momenta
\begin{equation}
\quad {\mathcal{H}}_{\mu }=\frac{1}{2}(p+i\mu )^{2}+\frac{1}{2}\sum\limits_{\alpha
\in \Delta }\hat{g}_{\alpha }^{2}V(\alpha \cdot q), \label{shift}
\end{equation}
and re-defined coupling constant
\begin{equation}
~~~\hat{g}_{\alpha
}^{2}=\left\{
\begin{array}{c}
g_{s}^{2}+\alpha _{s}^{2}\tilde{g}_{s}^{2}\quad \alpha \in \Delta
_{s} \\
g_{l}^{2}+\alpha _{l}^{2}\tilde{g}_{l}^{2}\quad \alpha \in \Delta
_{l} .
\end{array}
\right.
\end{equation}
Here $\Delta_{l}$ and $\Delta_{s}$ refer to the root system of the
long and short roots, respectively.

Thus we trivially have ${\mathcal{H}}_{\mu}=\eta ^{-1}h_{\text{Cal}}\eta$ with
$\eta =e^{-q\cdot \mu }$. Integrability follows then immediately by acting adjointly
with $\eta$ on the Calogero Lax pair $\dot{L}_{\text{Cal}}=\left[ L_{\text{Cal}},M_{\text{Cal}}\right] $,
such that the new pair is obtained by $L_\mu(p) = L_{\text{Cal}}(p+i\mu)$ and
$M_\mu = M_{\text{Cal}}$. An interesting statement is obtained by computing backwards
and allowing in (\ref{shift}) any kind of Calogero-Moser-Sutherland potential, i.e.
$V(x)=1/x^2$, $V(x)=1/sinh^2x$ or $V(x)=1/sin^2x$
\begin{equation}
{\mathcal{H}}_{\mu }=\frac{1}{2}p^{2}+\frac{1}{2}\sum\limits_{\alpha \in
\Delta }\hat{g}_{\alpha }^{2}V(\alpha \cdot q)+i\mu \cdot p-\frac{1}{2}\mu
^{2}. \label{int}
\end{equation}
By construction the Hamiltonian (\ref{int}) corresponds to an integrable model, but it
turns out \cite{AF} that the relation $\mu ^{2}=\alpha _{s}^{2}\tilde{g}_{s}^{2}\sum\nolimits_{\alpha \in \Delta
_{s}}V(\alpha \cdot q)+\alpha _{l}^{2}\tilde{g}_{l}^{2}\sum\nolimits_{\alpha
\in \Delta _{l}}V(\alpha \cdot q)$ is only valid for rational potentials. Thus without the
$\mu^2$-term only the deformed version of the Calogero model remains integrable and not
its generalizations.

\subsection{Complex deformed Calogero-Moser-Sutherland models}
Having seen that merely adding terms to complex Hamiltonians leads to rather
simple models, we comment on some of the other possibilities indicated in (\ref{fdef}),
which were explored in \cite{AFMZ}. One of the symmetries of the CMS-models
is its invariance with respect to the entire Coxeter group $\mathcal{W}$ resulting from the fact
that we sum over all roots and
the property that Weyl reflections preserve inner products. Interpreting now each Weyl reflection
as a parity transformation across a particular hyperplane, we may try to seek models
which remain invariant with regard to the action of a across these hyperplanes deformed version 
of the Weyl group $\mathcal{W}^{\mathcal{PT}}$ associated with some newly defined complex roots ${\tilde{\alpha}}$
\begin{equation}
\quad {\mathcal{H}}_{\mathcal{PT}\text{CMS}}=\frac{p^{2}}{2}+\frac{m^{2}}{16}
\sum\limits_{\tilde{\alpha}\in \tilde{\Delta}_{s}}(\tilde{\alpha}\cdot q)^{2}+\frac{1}{2}%
\sum\limits_{\tilde{\alpha}\in \tilde{\Delta}}g_{\tilde{\alpha}}V(\tilde{%
\alpha}\cdot q), \label{PTCMS}
\end{equation}
where $m,g_{\tilde{\alpha}}\in {\mathbb{R}}$. We outline the main features
of the construction of the complex root system
$\tilde{\alpha} \in {\mathbb{R}}^{n}\oplus i{\mathbb{R}}^{n}$ with the desired features.
First recall that to each simple root $\alpha _{i}$ there is an associated Weyl reflections
$ \sigma _{i}(x)=x- 2 \alpha_{i} (x\cdot \alpha _{i})/(\alpha _{i}^{2})$. The aim is
then to construct a new complex root system $\tilde{\Delta}$ in one-to-one correspondence
to the standard one $\Delta $, which may be recovered in the limit $\epsilon=\varepsilon-1 \rightarrow 0$
\begin{equation}
\lim_{\epsilon \rightarrow 0}\tilde{\alpha}_{i}(\epsilon )=\alpha
_{i}\qquad \text{for }\tilde{\alpha}_{i}(\epsilon )\in \tilde{\Delta}
(\epsilon ),\alpha _{i}\in \Delta .
\end{equation}
We define a ${\mathcal{PT}}$-Weyl reflection as $\tilde{\sigma}_{\alpha _{i}}:=\sigma _{\alpha _{i}}{\mathcal{T}}$,
where the time reversal ${\mathcal{T}}$ has the effect of a complex conjugation. We may then
use the action on a generic complex root $\tilde{\alpha}$ to determine their form
\begin{eqnarray}
\tilde{\sigma}_{\alpha _{j}}\left( \tilde{\alpha}_{j}(\epsilon )\right)
&=&\sigma _{\alpha _{j}}{\mathcal{T}}\left( {\texttt{Re}}\tilde{\alpha}%
_{j}(\epsilon )\right) +\sigma _{\alpha _{j}}{\mathcal{T}}\left( i {\texttt{Im}}%
\tilde{\alpha}_{j}(\varepsilon )\right)  \\
&=&\sigma _{\alpha _{j}}\left( {\texttt{Re}}\tilde{\alpha}_{j}(\epsilon
)\right) -i\sigma _{\alpha _{j}}\left( {\texttt{Im}}\tilde{\alpha}
_{j}(\epsilon )\right) \\
&=&- {\texttt{Re}}\tilde{\alpha}_{j}(\epsilon )-i {\texttt{Im}}\tilde{\alpha}
_{j}(\epsilon ) \\
&=&-\tilde{\alpha}_{j}(\epsilon).
\end{eqnarray}
As a solution to these equations we find
\begin{eqnarray}
\quad {\texttt{Re}}\tilde{\alpha}_{i}(\epsilon )&=&R(\epsilon )\alpha _{i}\text{%
\qquad and\qquad } {\texttt{Im}}\tilde{\alpha}_{i}(\epsilon )=I(\epsilon
)\sum\limits_{j\neq i}\kappa _{j}\lambda _{j}, \\
\lim_{\epsilon \rightarrow 0}R(\epsilon )&=&1\text{\qquad \qquad and\qquad }%
\lim_{\epsilon \rightarrow 0}I(\epsilon )=0,
\end{eqnarray}
with $\lambda_j$ denoting fundamental roots and $\kappa_j \in {\mathbb{R}}$.
Concrete examples for some specific algebras are for instance the deformed roots for $A_2$
\begin{eqnarray}
\tilde{\sigma}_{1}\tilde{\alpha}_{1}(\epsilon )&=&-R(\epsilon )\alpha
_{1}\mp iI(\varepsilon )\lambda _{2}=:-\tilde{\alpha}_{1}(\varepsilon ),  \\
-\tilde{\sigma}_{1}\tilde{\sigma}_{2}\tilde{\sigma}_{1}\tilde{\alpha}
_{1}(\varepsilon )&=&R(\varepsilon )\alpha _{2}\mp iI(\varepsilon )\lambda
_{1}=:\tilde{\alpha}_{2}(\varepsilon ),
\end{eqnarray}
or those for $G_2$
\begin{eqnarray}
\tilde{\alpha}_{1}(\epsilon ) &=&R(\epsilon )\alpha _{1}\pm
iI(\epsilon )\lambda _{2}  ,\\
\tilde{\alpha}_{2}(\epsilon ) &=&R(\epsilon )\alpha _{2}\mp
i3I(\epsilon )\lambda _{1}.
\end{eqnarray}
\par Having assembled the mathematical tools, we may substitute the deformed
roots into the model (\ref{PTCMS}) and study its properties. In \cite{AFMZ}
it was found that the $A_2$ and $G_2$ deformed Calogero models can still be solved by
separation of variables, analogously to the undeformed case. However, some
of the physical properties change, most notably the energy spectrum is
different. The reason for this difference is that some restrictions cease
to exist. For instance the wavefunctions are now regularized, such that
no restriction arises from demanding finiteness. The original
energy spectrum $ E=2\left\vert \omega \right\vert (2n+\lambda +1)$
becomes in the deformed case \cite{AFMZ}
\begin{equation}
E_{n\ell }^{\pm }=2|\omega |\left[ 2n+6(\kappa _{s}^{\pm }+\kappa _{l}^{\pm
}+\ell )+1\right] \qquad \text{for }n,\ell \in {\mathbb{N}}_{0},
\end{equation}
with $\kappa _{s/l}^{\pm }=(1\pm \sqrt{1+4g_{s/l}})/4$ and $s$, $l$ referring to 
coupling constants multiplying terms involving short
and long roots, respectively.
\par In section 2 our starting point were real fields and we naturally ended up
with complex particle systems, whereas in this section we started directly from the latter.
It remains to comment on how the relation may be established in the reverse procedure.
Having constructed the deformed roots we may compute the dual canonical coordinates $\tilde{q}$ from
\begin{equation}
\tilde{\alpha} \cdot q = \tilde{q} \cdot \alpha, \qquad \alpha ,q\in {\mathbb{R}}\text{, }%
\tilde{\alpha},\tilde{q}\in {\mathbb{R}}\oplus i{\mathbb{R}},
\end{equation}
and subsequently simply replace in (\ref{BK}) the Hamiltonian
${\mathcal{H}}_{\mathcal{PT}\text{CMS}}(p,q,\tilde{\alpha})$ by
${\mathcal{H}}_{{\mathcal{PT}}\text{CMS}}(\tilde{p},\tilde{q},\alpha)$.
The freedom in the choice of the functions $R(\epsilon),I(\epsilon)$
may then be used to satisfy the constraint $\texttt{grad} I=0$.
The dual canonical coordinates for $A_2$ are for instance easily computed to
\begin{eqnarray}
\tilde{q}_1 &=& R(\varepsilon)   q_1+ i I(\varepsilon)/3 (q_2 - q_3), \nonumber\\
\tilde{q}_2 &=&  R(\varepsilon) q_2 + i I(\varepsilon)/3 (q_3 - q_1), \\
\tilde{q}_3 &=&  R(\varepsilon)  q_3  + i I(\varepsilon)/3 (q_1 - q_2) \nonumber .
\end{eqnarray}
At this point it might still not be possible to satisfy $\texttt{grad} I=0$. As the construction
outlined is by no means unique one still has the additional freedom to employ 
an alternative one. These issues are further elaborated on in \cite{PAM}.

\section{Complex field equations}
Naturally we may also start directly by considering complex field
equations.

\subsection{${\mathcal{PT}}$-deformed field equations}
Taking the symmetries of the Korteweg deVries (KdV) equation into account one
may deform the derivatives according to (\ref{fdef}) either in
the second or the third term
\begin{equation}
u_{t}-6uu_{x ;\varepsilon}+u_{xxx ;\mu }=0,
\qquad \qquad \qquad  \varepsilon ,\mu \in {\mathbb{R}}. \label{KDVPT}
\end{equation}
The first possibility, i.e. $\mu=1$, was investigated in \cite{BendKdV},
leading to a not Galilean invariant, non-Hamiltonian system
with at least two conserved charges in form of infinite sums
allowing for steady wave solutions. Shortly afterwards the
second option was investigated in \cite{AFKdV}, that is $\varepsilon=1$,
giving rise to Galilean invariant Hamiltonian system with at
least three simple charges and steady state solutions.
The question of whether these systems are integrable
was thereafter addressed in \cite{Pain} by carrying out the Painlev\'{e}
test.

\subsection{The Painlev\'{e} test} \label{Ptest}
Let briefly summarize the main steps of this analysis proposed originally
in \cite{WTC}. The starting point is a series expansion for the field
\begin{equation}
u(x,t)=\sum\limits_{k=0}^{\infty }\lambda _{k}(x,t)\phi (x,t)^{k+\alpha }, \label{Pexp}
\end{equation}
with $\alpha$ being the leading order singularity in the field equation and
$\lambda _{k}(x,t)$ and $\phi (x,t)$ some newly introduced fields. The
substitution of this so-called Painlev\'{e} expansion into the partial differential
equation (PDE) under investigation leads to recurrence relations of the general form
\begin{equation}
g(j,\phi _{t},\phi _{x},\phi _{xx},\ldots )\lambda _{j}=f(\lambda
_{j-1},\lambda _{j-2},\ldots ,\lambda _{1},\lambda _{0},\phi _{t},\phi
_{x},\phi _{xx},\ldots ), \label{rec}
\end{equation}
with $f$ and $g$ being some functions depending on the individual system under consideration.
Solving these equations recursively might then lead at some level, say $k$,
to $g=0$. For that level we may then compute the right hand side of (\ref{rec})
and find that either $f\neq0$ or $f=0$. In the former case the Painlev\'{e}
test fails and the equation under investigation is not integrable, whereas in
the latter case $\lambda_k$ is found to be a free parameter, a so-called resonance. In case
the number of resonances equals the order of the PDE
the Painlev\'{e} test is passed, since in that scenario the expansion (\ref{Pexp})
has enough free parameters to accommodate all possible initial conditions.
A slightly stronger statement is made when the series is also shown to converge.
In that case one speaks of the Painlev\'{e} property of the PDE, which is conjectured
to be equivalent to integrability.

\subsection{Painlev\'{e} test for Burgers equation}
Instead of deformed KdV-equation (\ref{KDVPT}) let comment first on a simpler $\cal{PT}$-symmetrically deformed model, i.e.
Burgers equations
\begin{equation}
\quad u_{t}+uu_{x;\varepsilon }=i\kappa u_{xx;\mu }\text{\qquad \qquad with \ }%
\kappa ,\varepsilon ,\mu \in \mathbb{R}, \label{defB}
\end{equation}
for which the Painlev\'{e} test was carried out in \cite{Pain}.
In there it was found that leading order singularities only cancel for $\alpha=-1$
and $\varepsilon=\mu$ when taken to be integers. Keeping the $\varepsilon$ generic and starting with the
lowest order, the recurrence relations lead to the following equations
\begin{equation}
\quad
\begin{array}{lr}
\text{order }-(2\varepsilon +1)\text{: \ \  } & \lambda
_{0}+i2\varepsilon \kappa \phi _{x}=0, \\
\text{order }-2\varepsilon \text{:} & \phi _{t}\delta _{\varepsilon
,1}+\lambda _{1}\phi _{x}-i\kappa \varepsilon \phi _{xx}=0, \\
\text{order }-(2\varepsilon -1)\text{:} & \qquad \partial _{x}(\phi
_{t}\delta _{\varepsilon ,1}+\lambda _{1}\phi _{x}-i\kappa \varepsilon \phi
_{xx})=0.
\end{array}
\end{equation}
This means that
\begin{equation}
\quad \lambda _{0}=-i2\varepsilon \kappa \phi _{x}, \quad \lambda
_{1}=(i\varepsilon \kappa \phi _{xx}-\phi _{t}\delta _{\varepsilon ,1})/\phi
_{x}, \quad   \lambda _{2}\equiv\text{arbitrary}, \label{lam}
\end{equation}
such that we have already one of the desired free parameters. A more generic
argument can be used \cite{Pain} to derive a necessary condition for a resonance
to exist
\begin{equation}
i2^{\varepsilon -1}\varepsilon ^{\varepsilon } \lambda_r (r+1)(r-2)\kappa
^{\varepsilon }\phi _{x}^{2\varepsilon }=0. \label{free}
\end{equation}
The requirement is that the parameter $\lambda_r$ becomes free, which for
(\ref{free}) is obviously the case when $r=-1,2$. The values $2$ was already found in
(\ref{lam}) and $-1$ corresponds to the so-called fundamental resonance, which
seems to be always present. Thus according to the strategy outlined in section
\ref{Ptest}, we have the desired amount of free parameters and conclude that
the deformed Burgers equation (\ref{defB}) with $\varepsilon=\mu$ passes the
Painlev\'{e} test.
\par For the case $\varepsilon=2$ the convergence of the series was proven in \cite{Pain}
and concluded that the system even possess the Painlev\'{e} property and is therefore
integrable. In that case the expansion becomes
\begin{equation}
\quad u(x,t)=-\frac{4i\kappa }{\phi }+\lambda _{2}\phi +\frac{\xi ^{\prime }}{%
8\kappa }\phi ^{2}-\frac{i\lambda _{2}^{2}}{20\kappa }\phi ^{3}-\frac{%
i\lambda _{2}\xi ^{\prime }}{96\kappa ^{2}}\phi ^{4}+{\mathcal{O}}(\phi ^{5}).
\end{equation}
One may even find a closed solution when making the further assumption
$u(x,t)=\zeta (z)=\zeta (x-vt)$ and thus reducing the PDE to an ODE
\begin{equation}
\zeta (z)=e^{i\pi 5/3}(2v\kappa )^{1/3}\frac{\tilde{c}~Ai^{\prime }(\chi
)+Bi^{\prime }(\chi )}{\tilde{c}~Ai(\chi )+Bi(\chi )}
\end{equation}
with $\chi =e^{i\pi /6}(vz-c)(2v\kappa)^{-2/3}$ and $Ai$, $Bi$ denoting Airy functions.
\par
For the deformed KdV-equation (\ref{KDVPT}) is was concluded that they are only integrable
when $\varepsilon=\mu$, albeit the Painlev\'{e} series was found to be defective meaning
that is does not contain enough resonances to match the order of the PDE.

\subsection{Compactons versus Solitons}
A further interesting $\cal{PT}$-symmetric deformation was proposed \cite{PTcomp}
for the generalized KdV equations \cite{Comp}, which are known to possess compacton solutions
\begin{equation}
{\mathcal{H}}_{l,m,p}=-\frac{u^{l}}{l(l-1)}-\frac{g}{m-1}u^{p}(iu_{x})^{m} . \label{comppt}
\end{equation}
The corresponding equations of motions are
\begin{eqnarray}
&&u_{t}+u^{l-2}u_{x}+gi^{m}u^{p-2}u_{x}^{m-3}\left[ p(p-1)u_{x}^{4}\right.
\quad  \nonumber \\
&&\left. +2pmuu_{x}^{2}u_{xx}+m(m-2)u^{2}u_{xx}^{2}+mu^{2}u_{x}u_{xxx}\right]
=0.
\end{eqnarray}
The system (\ref{comppt}) is yet another generalization of the generalized KdV system
and the $\cal{PT}$-symmetric deformation of the KdV equation suggested in \cite{AFKdV},
i.e. $\mu=1$ in (\ref{KDVPT}), corresponding to ${\mathcal{H}}_{l,2,p}$ and
${\mathcal{H}}_{3,\varepsilon+1,0} $, respectively.
These models were found to admit compacton solutions, which, depending on $l,p$ or $m$,
are either unstable, stable with independent width $A$ and amplitude $\beta$ or
stable with amplitudes $\beta$ depending on the width $A$. An interesting question
to ask is whether these systems also allow for soliton solutions, which would be the case
when they pass the Painlev\'{e} test. The test was carried out in \cite{compact} and the results
are summarized in the following table:
\begin{center}
\begin{tabular}{|l||l|l|}
\hline
${\mathcal{H}}_{l,m,p}$ & compactons & solitons \\ \hline\hline
$l=p+m$ & stable, independent $A,\beta $ & no \\ \hline
$2<l<p+3m$ & stable, dependent $A,\beta $ & yes \\ \hline
$l\leq 2$ or $l\geq p+3m$ & unstable & no \\ \hline
\end{tabular}
\end{center}
Interestingly, it was found that there is no distinction between the
generalized KdV equations and $\cal{PT}$-symmetric extensions of
generalized KdV equations with regard to the Painlev\'{e} test.

\section{Conclusions}

We have shown that $\cal{PT}$-symmetry can be used as a guiding principle to define new interesting models, 
some of which are even integrable. Some complex particle systems, possibly restricted to some submanifold, 
are shown to be equivalent to some real valued fields obeying non-linear field equations. We also investigated 
some complex field equations, which for certain choices of the parameters involved turn out to be integrable
admitting soliton as well as compacton solutions.

\medskip

\noindent \textbf{Acknowledgments}. I would like to thank Sudhir Jain and Zafar Ahmed,
for all their efforts to organize PHHQP VIII and their kind hospitality. I am grateful to Paulo Assis, Bijan Bagchi, Carla
Figueira de Morisson Faria and Miloslav Znojil for collaboration.

\end{document}